# Stochastic Variational Bayesian Inference for a Nonlinear Forward Model

Michael A. Chappell*[1,2,3], Martin Craig[1,2,3], Mark W. Woolrich[4]

[1] Sir Peter Mansfield Imaging Centre, School of Medicine, University of Nottingham, UK.
[2] Institute of Biomedical Engineering, Department of Engineering Science, University of Oxford, UK.
[3] Wellcome Centre for Integrative Neuroimaging, FMRIB, Nuffield Department of Clinical Neurosciences, University of Oxford, UK.
[4] Oxford Centre for Human Brain Activity, Wellcome Centre for Integrative Neuroimaging, Department of Psychiatry, University of Oxford, UK.
* email: michael.chappell@nottingham.ac.uk

*Abstract*—Variational Bayes (VB) has been used to facilitate the calculation of the posterior distribution in the context of Bayesian inference of the parameters of nonlinear models from data. Previously an analytical formulation of VB has been derived for nonlinear model inference on data with additive gaussian noise as an alternative to nonlinear least squares. Here a stochastic solution is derived that avoids some of the approximations required of the analytical formulation, offering a solution that can be more flexibly deployed for nonlinear model inference problems. The stochastic VB solution was used for inference on a biexponential toy case and the algorithmic parameter space explored, before being deployed on real data from a magnetic resonance imaging study of perfusion. The new method was found to achieve comparable parameter recovery to the analytic solution and be competitive in terms of computational speed despite being reliant on sampling.

*Index Terms*— Biomedical signal processing, Bayesian inference, magnetic resonance imaging, nonlinear estimation, signal processing.

## I. Introduction

BAYESIAN methods have proved popular in a wide range of data analysis applications for the inference of parameters in the presence of noise from both well-specified deterministic mathematical models of systems, as well as in data-driven machine learning applications. One of the appeals of Bayesian inference methods is that they take a principled approach to incorporating prior information and the ability to derive measures of confidence, or uncertainty, in the estimated parameter values. However, few practical analytical algorithms exist for Bayesian inference applications, apart from trivial problems, because the integrals involved in finding the posterior distribution, or summary statistics of the posterior distribution, are often intractable.

A range of approximate or sampling-based solutions have been developed [1]. Notably Variational Bayesian (VB) methods have been developed that approximate the true posterior distribution by means of another tractable distribution (or product of distributions), recasting the inference problems as one of achieving the best match between approximate and true distribution using the information-theoretic Kullback Leibler (KL) divergence measure [2]. VB solutions are typically less computationally demanding than alternative approaches that involve sampling the posterior distribution and have thus become popular in data analysis methods. However, the original implementation of the VB method requires tractable integrals in the evaluation of the KL-divergence, restricting the combination of problems and approximating distributions that can be used [3]. More recently stochastic VB inference schemes have been proposed which adopt the same approximate posterior strategy, but perform the optimization using stochastic gradient descent [4].

A particularly common class of data analysis problem for which Bayesian inference can be beneficial is that of inferring parameters of a non-linear mathematical model from noisy data. In the context of this work, a good example of this problem being the analysis of functional MRI data, where a series of images with different contrasts are exploited to make measurements of some underlying process. Notably, the relationship between the parameters of interest and the measured imaging data in each image element (voxel) can be described in terms of a non-linear mathematical model. This type of application involves individual model fitting in tens to hundreds of thousands of voxels for one dataset, using data that typically has a poor signal-to-noise ratio (SNR). The availability of a mathematical model and the value of including prior information to help regularize the output in the face of poor SNR, has motivated the use of Bayesian inference for this type of data. Previously, a framework was developed for This work was supported in part by the Engineering and Physical Sciences Research Council UK (EP/P012361/1). The Wellcome Centre for Integrative Neuroimaging is supported by core funding from the Wellcome Trust (203139/Z/16/Z). MWW's research is supported by the NIHR Oxford Health Biomedical Research Centre, by the Wellcome Trust (106183/Z/14/Z), and the MRC UK MEG Partnership Grant (MR/K005464/1).

inference of the parameters of a non-linear forward model from serial data using VB [5], where at each iteration of the algorithm a Taylor series approximation was used to linearize the model to make it tractable for classical VB. This solution has subsequently been applied in a range of medical imaging applications [6] [7] [8].

In this work we explore a stochastic Variational Bayes solution to the non-linear model inference problem. Showing that it has more flexibility in terms of prior specification than the existing VB method, and that solutions can be achieved on a realistic timescale by appropriate choice of the parameters that determine the optimization process.

## II. THEORY

### A. Bayesian Inference

If we have a series of measurements, $y$, and we wish to use them to determine the parameters, $w$, of our chosen model $M$, Bayesian inference frames this problem as

$$p(w|y,M) = \frac{p(y,w|M)}{p(y|M)} = \frac{p(y|w,M)p(w|M)}{p(y|M)} \quad (1)$$

which gives the *posterior* probability of the parameters given the data and the model, $p(w|y,M)$, in terms of: the *likelihood* of the data given the model, $p(y|w,M)$, the *prior* probability of the parameters for this model, $p(w|M)$, and the *evidence* for the measurements given the chosen model, $p(y|M)$.

In practice, we often desire a best estimate of the parameter(s), $w$, from the posterior distribution, and some measure of uncertainty. A suitable solution is to take moments of the posterior distribution, e.g., the mean and the variance. This requires integration in order to normalise or marginalise the posterior distributions and is generally where challenges arise: the posterior distribution is often intractable.

### B. Variational Inference

Variational Bayes involves choosing a distribution to use as an *approximate posterior*, $q(\theta)$, and finding the version of it that is as close as possible to the true posterior [3]. To measure 'closeness' the KL-divergence between the two distributions is used; in practice, this is equivalent to maximizing the (negative variational) free energy (also called the Evidence Lower Bound, or ELBO)

$$F(\theta) = \int q(\theta) \log\left(p(y|\theta)\frac{p(\theta)}{q(\theta)}\right) d\theta \quad (2)$$

One option is to choose a parameterized form for the approximate posterior: $q(\theta;\zeta)$, where $\zeta$ is the set of hyper-parameters. We can then appeal to the calculus of variations to derive a series of 'update equations' for the hyper-parameters. Since this also involves an integration, this method places a number of constraints on the choice of the approximate posterior distribution; for example, the use of conjugate distributions [3].

An alternative is a 'brute force' approach, maximising $F$ directly using gradient descent, this requires the gradients of $F$ with respect to the hyper-parameters $\zeta$

$$\nabla_\zeta F(\theta) = \nabla_\zeta \left(\int q(\theta) \log\left(p(y|\theta)\frac{p(\theta)}{q(\theta)}\right) d\theta\right) \quad (3)$$

This will generally not be tractable in practice, but sampling can be used, i.e., taking a Monte Carlo approximation, for the required integral

$$F \approx \frac{1}{L}\sum_{l=1}^{L} \log(p(y|\theta^{*l})) - \log\left(\frac{q(\theta^{*l})}{p(\theta^{*l})}\right) \quad (4)$$

Where $\theta^{*l}$ are drawn from $q(\theta)$. Thus, we can write the gradient as

$$\nabla_\zeta F \approx \frac{1}{L}\sum_{l=1}^{L} \nabla_\zeta \left(\log(p(y|\theta^{*l})) - \log\left(\frac{q(\theta^{*l})}{p(\theta^{*l})}\right)\right) \quad (5)$$

Minimization of the free energy can now be cast in the form of a stochastic gradient descent (sGD) problem. This approach requires the analytical computation of the gradients. For many analytical forward models, this can be achieved using Automatic Differentiation; for example, using backpropagation.

For this formulation, convergence might be an issue since we are working with 'noisy' estimates of the gradient. This may actually be advantageous for avoiding local minima in non-linear model fitting where convergence to a global minimum is not guaranteed, since using stochastic estimates will naturally cause wider exploration of the parameter space. However, it might lead to slow convergence as the gradient estimates are highly variable, since the samples we take to calculate the gradient depend upon the current estimate of $\zeta$.

A method that has been adopted by the machine learning community to reduce variability for sGD is to use a pathwise gradient estimator (also known as the 'reparameterization trick') [9] where random samples are drawn from a distribution independent of $\zeta$ and used to calculate an estimate of the required gradient via a transformation. For example, we may sample from a standard normal distribution $\epsilon \sim N(0,1)$ and transform it (reparametrize) into a sample from $q(\theta)$

$$\theta^* = F_\zeta^{-1}(\Phi(\epsilon)) \quad (6)$$

where $\Phi(\epsilon)$ is the cumulative distribution function for the standard normal and $F_\zeta$ is the CDF for $q(\theta)$.

A final additional option to improve computational efficiency, where the forward model calculations are a bottleneck to computation, is to use 'mini-batches'. This involves computing the gradients using only a subset of the data points at each iteration of the sGD algorithm. Typically, an implementation would split the data into multiple batches and run sequentially though them.

### C. Inference for a Non-Linear Forward Model

The problem of inferring parameters from series data using a non-linear model can be written as

$$\mathbf{y} = \mathbf{g}(\boldsymbol{\theta}) + \mathbf{e} \quad (7)$$

where $\mathbf{y}$ is a vector of measured data of length $N$, with $N$ being the number of samples, $\mathbf{g}$ is a non-linear model with parameters $\theta$, and $\mathbf{e}$ is vector of noise contributions of length $N$. If we assume white noise of the form $\mathbf{e} \sim N(\mathbf{0}, \phi\mathbf{I})$, then for $N$ observations the log-likelihood is

$$\log P(\mathbf{y}|\Theta) = \frac{N}{2}\log\phi$$
$$-\frac{1}{2}\phi(\mathbf{y}-\mathbf{g}(\boldsymbol{\theta}))^{\mathrm{T}}(\mathbf{y}-\mathbf{g}(\boldsymbol{\theta})) \quad (8)$$

where $\Theta = \{\boldsymbol{\theta}, \phi\}$ is the set of all model and noise parameters.

*D. Analytical solution for VB inference*

For the analytical Variational Bayes (aVB) scheme in [5] the approximate posterior was factorized and separate conjugate prior-posterior pairs were chosen for the model and noise parameters based on the likelihood, namely

$$q(\Theta) = q(\boldsymbol{\theta})q(\phi) = MVN(\boldsymbol{\theta};\boldsymbol{\mu},\boldsymbol{\Lambda}^{-1})\,\mathrm{Gamma}(\phi;s,c)$$
$$P(\Theta) = P(\boldsymbol{\theta})P(\phi) = MVN(\boldsymbol{\theta};\boldsymbol{\mu}_0,\boldsymbol{\Lambda}_0^{-1})\,\mathrm{Gamma}(\phi;s_0,c_0) \quad (9)$$

where $MVN(\boldsymbol{\theta};\boldsymbol{\mu},\boldsymbol{\Lambda}^{-1})$ is a multi-variate normal distribution with vector of mean values $\boldsymbol{\mu}$ and precision matrix $\boldsymbol{\Lambda}$, and $\mathrm{Gamma}(\phi;s,c)$ is a gamma distribution with shape and scale parameters $s$ and $c$ respectively.

The application of the calculus of variations results in a set of update equations for each of the hyperparameters ($\boldsymbol{\mu}, \boldsymbol{\Lambda}, s, c$). To deal with the non-linearities in the model a local linear approximation using a first-order Taylor series expansion

$$\mathbf{g}(\boldsymbol{\theta}) \approx \mathbf{g}(\boldsymbol{\mu}) + \mathbf{J}(\boldsymbol{\theta} - \boldsymbol{\mu}) \quad (10)$$

with

$$J_{j,k} = \left.\frac{\mathrm{d}g(\theta_j)}{\mathrm{d}\theta_k}\right|_{\boldsymbol{\theta}=\mathbf{m}} \quad (11)$$

The application to a non-linear forward-model means that convergence to a global minimum is not guaranteed. Convergence was monitored using the free energy calculated at each iteration. In [5] a 'trial' mode of convergence detection was proposed, whereby the algorithm was run for a fixed number of iterations, but if there was a reversal of the free energy before the maximum iterations had been reached a further fixed number of trial iterations were attempted. If an improvement in free energy was achieved the iterations continued until the maximum, otherwise the results prior to free energy reversal was returned.

## III. METHODS

*A. Implementation*

Stochastic variational inference is not restricted to conjugate priors; in this work we use a MVN posterior distribution over all the parameters (including that of the noise distribution):

$$q(\Theta) = q\left(\begin{bmatrix}\boldsymbol{\theta}\\-\log(\phi)\end{bmatrix}\right) = MVN(\boldsymbol{\theta};\mathbf{m},\mathbf{C}) \quad (12)$$

where we use a log-scale for the noise parameter to constrain it to positive values. By choosing this formulation we allow for covariance between noise and model parameters, something not possible under the original aVB derivation. Additionally, this approximate posterior is amenable to the use of reparameterization gradients. Namely, we can draw samples using

$$\Theta^* = \mathbf{m} + \mathbf{S}\boldsymbol{\epsilon}^* \quad (13)$$

with $\boldsymbol{\epsilon} \sim MVN(\mathbf{0},\mathbf{I})$, and $\mathbf{S}$ is the Cholesky decomposition of $\mathbf{C}$, i.e., $\mathbf{C} = \mathbf{S}\mathbf{S}^{\mathrm{T}}$.

There is no need for conjugacy between prior and approximate posterior. Although, for this work we chose to use a MVN prior over all parameters (which is close to the conjugate prior used in the original aVB derivation [5])

$$P(\Theta) = P\left(\begin{bmatrix}\boldsymbol{\theta}\\-\log(\phi)\end{bmatrix}\right) = MVN(\mathbf{m}_0,\mathbf{C}_0) \quad (14)$$

This allows us to write the KL-divergence between approximate posterior and prior, by combining (12) and (14), as

$$\log\left(\frac{q(\Theta)}{P(\Theta)}\right) = -\frac{1}{2}\log\left(\frac{|\mathbf{C}|}{|\mathbf{C}_0|}\right)$$
$$-\frac{1}{2}(\Theta-\mathbf{m})^{\mathrm{T}}\mathbf{C}^{-1}(\Theta-\mathbf{m})$$
$$+\frac{1}{2}(\Theta-\mathbf{m}_0)^{\mathrm{T}}\mathbf{C}_0^{-1}(\Theta-\mathbf{m}_0) \quad (15)$$

Thus, in this particular case, we do not need a stochastic approximation for all of the terms in (4). Instead we can reduce our reliance on stochastic approximations for all of the terms and write it as:

$$F \approx \mathrm{Loss} + \frac{1}{L}\sum_L \log(p(y|\theta^{*l})) \quad (16)$$

with

$$\mathrm{Loss} = \int q(\boldsymbol{\theta})\log\left(\frac{q(\boldsymbol{\theta})}{p(\boldsymbol{\theta})}\right)d\boldsymbol{\theta}$$
$$= \frac{1}{2}\left\{\mathrm{Trace}(\mathbf{C}_0^{-1}\mathbf{C}) - \log\left(\frac{|\mathbf{C}|}{|\mathbf{C}_0|}\right)\right.$$
$$- N$$
$$\left. + (\mathbf{m}-\mathbf{m}_0)^{\mathrm{T}}\mathbf{C}_0^{-1}(\mathbf{m}-\mathbf{m}_0)\right\} \quad (17)$$

The sVB algorithm was implemented in Python 3.6.7 using the Tensor Flow library 1.10.0 and the Adam optimizer [10], the loss function was defined as the negative free energy, $-F$, given in (16) (i.e., in practice the algorithm sought to minimize the non-negative free energy).

The aVB algorithm was implemented using the C++ code included as part of fabber within the FMRIB Software Library (www.fmrib.ox.ac.uk/fsl).

*B. Assessing Convergence with Simulated Data*

Based on the implementation of the sVB algorithm there are three parameters that need to be chosen that will affect whether and how rapidly it converges. Namely:
1) Number of posterior samples ($L$);
2) Batch size, $B$;
3) Learning rate of the optimizer, $\alpha$.

These parameter choices were explored for the example problem of fitting a biexponential model to simulated data corrupted by white noise, following the analysis in [5]. The model was defined as

$$M(t) = A_1 e^{-R_1 t} + A_2 e^{-R_2 t} \quad (18)$$

with $A_1 = A_2 = 10$ and $R_1 = 1, R_2 = 10$. Data were generated with $N = 10, 20, 50, 100$ time points evenly spaced between $t = 0$ and $t = 5$. White noise was added using random draws from a normal distribution with zero mean and standard deviation of 1.0 and for each case 1000 noisy realizations were generated. Added noise with standard deviation of 2.0, 5.0 and

10.0 was also explored. Since for the biexponential model there are always two equivalent solutions obtained by exchanging $A_1, R_1$ with $A_2, R_2$, the results were 'normalized' by setting $A_1, R_1$ after the inference to correspond to the slower rate estimated. Optimization was performed as a single global problem over all 1000 realizations for each case, defining the loss as the mean free energy over all the realizations.

Two inference scenarios were considered on the simulated data: inference of the full posterior covariance matrix, **C**, and a simpler version where **C** was assumed to be diagonal (i.e., only the posterior variances of the model parameters where inferred).

Algorithmic parameters were explored by varying:
1) Learning rate: $\alpha = 0.005, 0.01, 0.02, 0.05, 0.1, 0.25, 0.5$ for $L = 2, 5, 10, 20, 50, 100, 200$, with $B = N$ (i.e., no batching), to find the range of learning rate that achieved the lowest free energy value at convergence;
2) Sample size: $L = 1, 2, 5, 10, 20, 50, 100, 200$, to find the sample size that achieved the fastest convergence;
3) Batch size: $B = 5, 10, 20, 50, 100$ (larger batch sizes only being used where there were sufficient timepoints in the data), using the learning rate and sample size from experiment 1. Batches were generated from uniform strides through the time series, so each contained examples from the full range of timepoints.

The algorithm was in all cases run for 500 epochs, where an epoch represents a complete pass through the data, i.e., where the batch size is smaller than the data size a single epoch might include multiple iterations of the gradient descent. The choice of parameters was determined by comparing the run time for convergence to be achieved (within a set percentage of the best achieved free energy value) and the mean of best free energy value achieved across the 1000 realizations. Run time was used in place of number of training epochs as it is more representative of real-world application, and to account for the fact that batch processing uses multiple iterations of the optimization loop within one epoch.

### C. Initialization

Like any non-linear fitting algorithm, it is necessary to set initial values for the parameters to be estimated. It would be expected that convergence to the solution, and rate of convergence, might depend on the choice of initial values. Like the aVB algorithm, for the sVB method it is necessary to set initial values for the (approximate) posterior distribution, i.e., **m** and **C**. We considered a number of combinations of initial posterior and choice of prior distribution:
1) Prior distribution
   a) Informative: mean equal to the true values and standard deviation of 2.0 for all parameters (no covariance);
   b) Noninformative: mean equal to 1.0 for all parameters and standard deviation of $10^6$ (no covariance).
2) Initial posterior distribution
   a) True: initial posterior matching the prior, which means it was initialized with the true mean values for the parameters.
   b) Data: initial posterior with amplitudes estimated from the data by setting the means for these parameters based on half the maximum amplitude of the data and true decay rates. Standard deviations matched those of the prior.
   c) Wrong: initial posterior with mean values of 1.0 for both decay rates and 100 for the amplitudes, i.e., deliberately far from the true solution. Standard deviations matching those of the prior.
   d) Uninformed: initial posterior that matched the noninformative prior distribution, i.e., all parameters are treated equally and with a generic distribution with high uncertainty.

The influence of the initial posterior for both type of prior information was compared using the mean free energy value obtained at convergence over the 1000 realizations and the run time to convergence.

### D. Application

The aVB algorithm was also applied to kinetic model-fitting for perfusion estimation from real Arterial Spin Labelling MRI brain data. This is similar to the application of aVB in [5] and represents an example of non-linear model fitting that has been shown to benefit from the application of prior information within a Bayesian inference setting. Briefly, ASL MRI is a perfusion imaging technique that can supply time series measurements of the inflow of a bolus of magnetically labeled blood water into individual voxels of tissue [11]. In this work, we consider a form of ASL MRI that uses pseudo-continuous labelling (PCASL), wherein blood-water is magnetically labeled in the arteries as it passes through a plane, with labeling continuing for a specified label duration. An image is subsequently acquired after a post label delay (PLD) that contains signal from both brain tissue and inflowing labeled blood water. By subtracting this image from another 'control' image without labeling, an image of inflow is obtained. Repetition of this process, varying the PLD, generates timeseries data in every voxel, allowing perfusion and other haemodynamic parameters to be estimated from this data using the kinetic model:

$$\Delta M(t) = \begin{cases} 0 & t < \Delta t \\ 2M_{0a}fT_{1app}(1 - e^{-\frac{t-\Delta t}{T_{1app}}}) & \Delta t \leq t < \Delta t + \tau \\ 2M_{0a}fT_{1app}e^{-\frac{\Delta t}{T_{1b}}}e^{-\frac{t-\Delta t-\tau}{T_{1app}}}(1 - e^{-\frac{\tau}{T_{1app}}}) & \Delta t + \tau \leq t \end{cases}$$
(19)

where $1/T_{1app} = 1/T_1 + f/\lambda$, $f$ is the perfusion (in units of s$^{-1}$), $\Delta t$ is the time of arrival of labelled blood-water in the tissue (normally termed the arterial transit time, ATT), $T_1$ and $T_{1b}$ are the longitudinal relaxation time constants for tissue and blood water respectively, $\tau$ is the label duration, $t$ is the time since labelling began ($\tau + PLD$), and $M_{0a}$ is the equilbirum magetnization of arterial blood, which is normally obtained from a separate calibration image.

The test data was from a single individual in which ASL MRI had been acquired using pcASL labeling with label duration $\tau = 1.8 \, s$ (this data is used in [11] and can be found at

www.neuroimagingprimers.org). The dataset contained a total of 96 measurements (48 pairs of labeled and control images) which included 6 PLDS (0.25, 0.5, 0.75, 1.0, 1.25, 1.5) each repeated eight times. The data contains a full 3D volume image of the brain acquired as a series of 2D slices, later slices thus had a longer PLD than earlier ones (each being 45.2 ms later than the one immediately preceding), this was incorporated into the model for estimation. The sVB and aVB methods were used to estimate $f$ and $\Delta t$ subject to priors $P(f) \sim N(0, 10^6)$ and $P(\Delta t) \sim N(1.3, 1)$, implemented via the MVN prior specified for both sVB and aVB with zero off-diagonal terms in the covariance matrix. The noise parameter was subject to the prior with mean 1.0 and variance $10^{-6}$, with initial posterior variance determined from the variance of the individual data series.

The convergence properties were examined on this data using the same procedure as for the biexponential model, examining minimum free energy achieved and time to convergence across all the voxels identified as being within the brain based on an existing brain mask. The same parameters were used are for the simulated data experiments, except that batch sizes of $B = 5, 6, 9, 12, 18, 24, 48$ were tested. As previously, these were generated from uniform strides through the timeseries. The ASL data set used contained all the repeats at a single PLD grouped together, so this strategy ensured that samples from all PLDs were included in each batch. The comparison of batch sizes of 5 and 6 was intended to assess whether there was any advantage to aligning the batches directly with the repeats of each PLD.

IV. RESULTS

A. Assessing Convergence with Simulated Data

Fig. 1 shows the minimum mean free energy achieved as a function of learning rate and posterior sample size for the different datasets generated using the biexponential model. Fig. S1 shows the mean free energy plotted as a function of run time for example cases. Overall, the lowest mean free energy value was achieved with $\alpha$ in the range 0.25 to 0.05, with the minimum free energy achieved most consistently at or near 0.05 across the range of posterior sample sizes tested. Small posterior samples sizes and larger values of learning rate were not reliably able to achieve a free energy near to the minimum. Posterior sample sizes above 10 were generally able to achieve the same minimum free energy as larger sample size values. Inference without covariance terms was more tolerant to a range of algorithmic parameters than the problems that included posterior covariance. Similar results were seen for data with greater noise added (see Fig. S2 for results with N=20), except that with greater noise the influence of learning rate and posterior sample size on best free energy achieved was reduced. A learning rate of 0.05 was adopted for subsequent experiments.

Fig. 2 shows the run time to convergence as a function of posterior sample size when using $\alpha = 0.05$. Overall, larger sample size resulted in slower convergence, but very small values, $L < 5$, reversed this trend in some cases. The results support sample sizes of $L = 10 - 20$ for the larger number of time points in the data ($N = 50 - 100$), with there being a benefit of having more samples, $L = 20 - 50$, with fewer data points, $N < 20$.

Fig. 3 shows mean free energy at convergence as a function of learning rate for a range of batch sizes. There was an interaction between choice of batch size and learning rate needed to achieve minimum mean free energy at convergence. However, the previously identified value of learning rate, $\alpha = 0.05$, appeared to be tolerant to a range of batch sizes. Fig. 4 shows the time to convergence for a range of batch sizes (only $N = 50$ and $100$ shown). Time to convergence was reduced by the use of batches, with a batch size of $B = 10$ generally best.

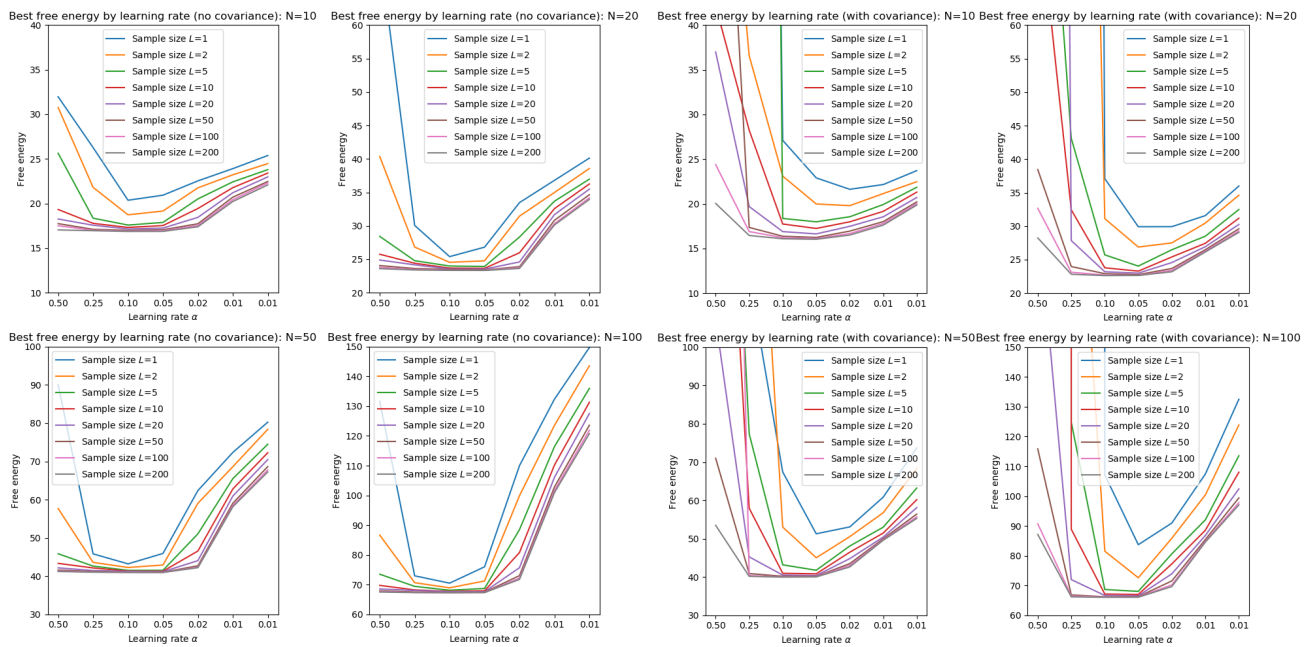

Fig. 1: Mean free energy achieved in the biexponential model fitting to simulated data experiments as learning rate and number of posterior samples used was varied. Across a range of data size (N) the optimal learning rate was around 0.05 and differences in mean free energy achieved were minimal for posterior samples size of 5 or more.

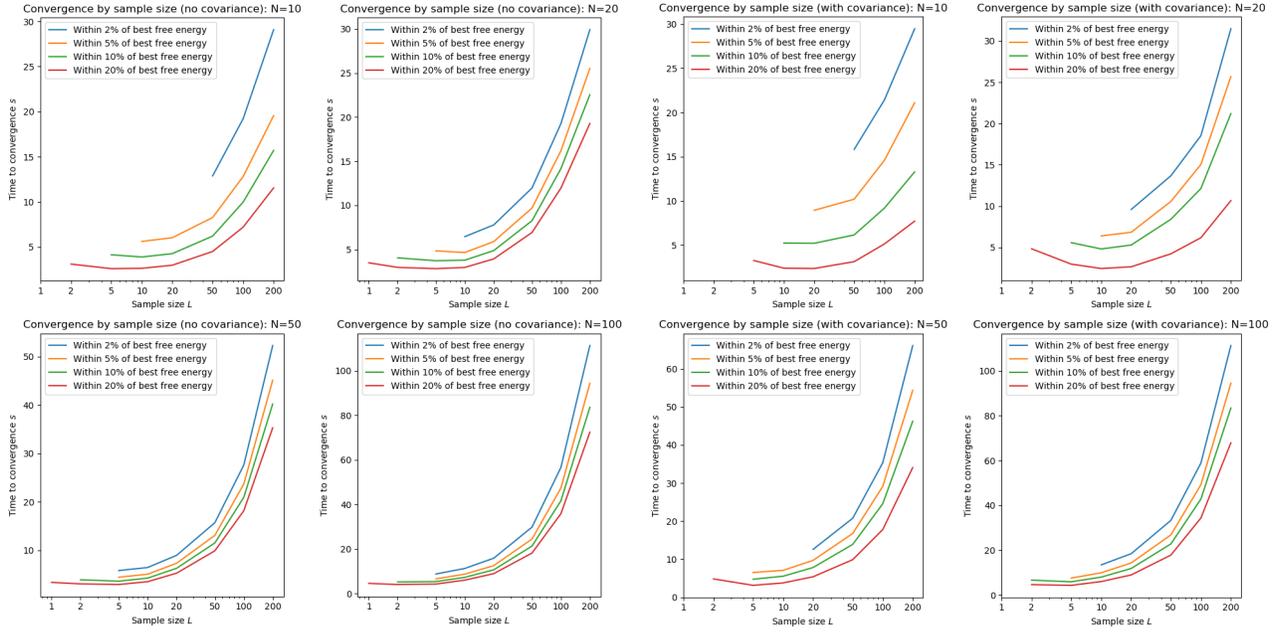

Fig. 2: Run time to convergence in the biexponential model fitting to simulated data experiments for a range of posterior sample size showing time to achieve a value within the specified tolerance of the minimum free energy achieved across all posterior sample sizes used, using a learning rate of 0.05. The shortest time to convergence was generally achieved using posterior sample size in the range 1-20.

### B. Comparison of sVB and aVB on simulated data

Fig. S3 shows box plots of the parameter estimates using both aVB and sVB algorithms on the simulated data, for sVB the best choice of learning rate, posterior sample and batch sizes were chosen from the convergence analysis ($\alpha = 0.05$, $L = 20$ and $B = 10$). The results for aVB and sVB were comparable, especially for larger N (50 or 100). For smaller N both algorithms tended to estimate a single exponential component.

### C. Choice of prior and initial posterior distributions on simulated data

Fig. 5 shows the minimum free energy achieved for $N = 100$

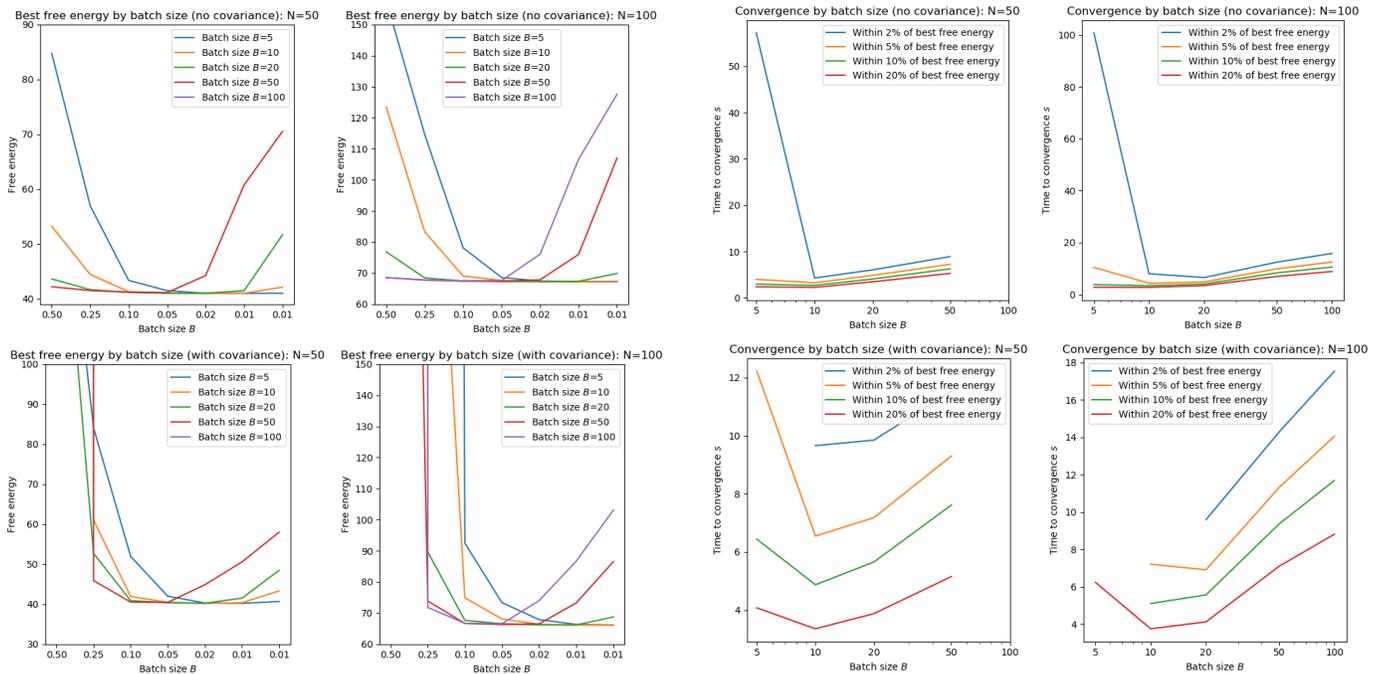

Fig. 3: Mean free energy at convergence in the biexponential model fitting to simulated data experiments as a function of learning rate and batch size with posterior sample size of 20. The same free energy value could be achieved at all batch sizes, with smaller batch sizes favoring smaller learning rate values.

Fig. 4: Run time to convergence in the biexponential model fitting to simulated data experiments as a function of batch size with posterior sample size of 20 and learning rate of 0.05. Batch sizes around 10 resulted in consistently shortest run time to convergence.

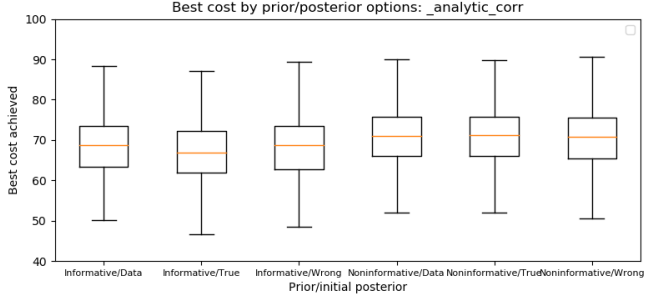

Fig. 5: Minimum free energy achieved for different combinations of prior and initial posterior distributions in the biexponential model fitting to simulated data experiments.

as the prior and initial posterior distribution was varied (not showing the cases where the initial posterior was non-informed, the full results can be found in Fig. S4). A small but significant difference was seen between the use of an informative and non-informative prior. Otherwise, the choice of initial posterior was not important as long as it was not non-informed (i.e., the standard deviation of the initial posterior was not set to be $\sim 10^6$). This was the case even if the initial estimate for the mean of the distribution was far from the true value. Where this was the case, convergence took a greater number of epochs (see Fig. S5).

### D. Application

Fig. 6 shows example perfusion and ATT images from the ASL MRI data from both sVB and aVB analysis, along with an example timeseries and model fit. Fig. 7 summarizes the convergence results for the ASL MRI dataset. Overall, the results were similar to the biexponential model and a learning rate of 0.05, posterior sample size of 5 and batch size of 12-18 resulted in the best convergence properties, with similar performance being found for values around those. Like the simulated data very small posterior sample size (i.e., L=2) was generally acceptable in comparison with the best results achieved.

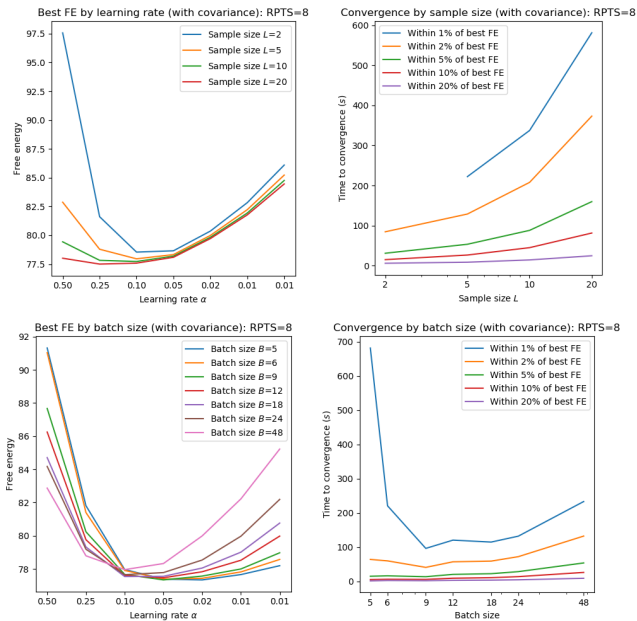

Fig. 7: Summary of the minimum mean free energy achieved with real ASL-MRI data and tie to convergence for the brain imaging data application for variation in learning rate, posterior sample size and batch size.

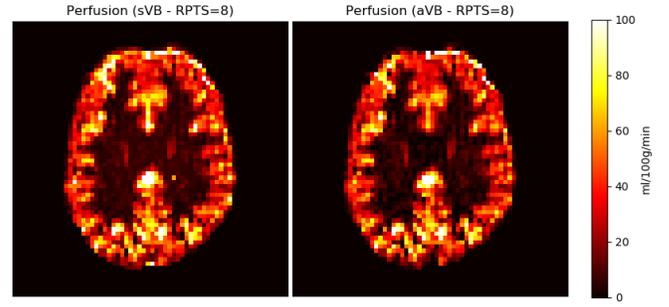

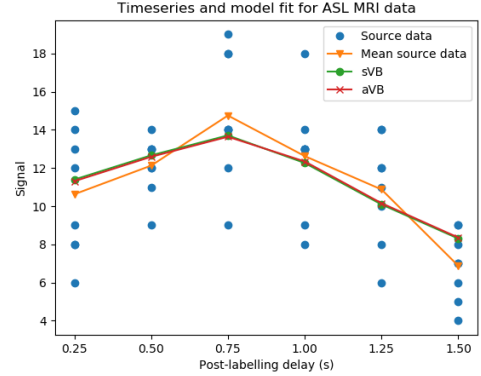

Fig. 6: ASL-MRI perfusion images after kinetic analysis using sVB and aVB algorithms, representative slices shown, and example fit to series data in a voxel.

## V. Discussion

In this work we have demonstrated the implementation of a stochastic solution for Variational Bayesian inference applied to non-linear model fitting. We have shown that the sVB algorithm can achieve very similar results to the existing aVB solution and have explored the influence of the algorithmic parameters on convergence. Unlike the analytic solution, sVB does not require an analytical solution to various integrals needed to arrive at an iterative scheme, this removes the restriction in the original aVB algorithm for conjugacy between prior and posterior distributions. Ultimately, this allows the sVB approach to be applied to a much wider range of problems including different noise models and priors. This has been exploited here for non-linear model fitting by using an MVN prior over all parameters (including the noise magnitude parameter), and avoids the need to make a Taylor approximation to the non-linear function. Although not exploited here, the aVB can trivially be implemented on GPU hardware offering substantial improvements in computation time for applications like the MRI parameter estimation one used here.

The trade-off in adopting a stochastic solution is that the algorithm proceeds using a number of 'noisy' samples which might hamper or even prevent it reaching the global optimum. In contrast, for a linear model the VB formulation guarantees convergence to the global minimum; although, this is not true for non-linear models and convergence was identified as an issue for the aVB algorithm [5]. In practice, a stochastic solution can assist in the avoidance of local minima because each step taken has a stochastic element that means there is always a probability of stepping out of a local minimum. A result of this is that it is harder to monitor convergence, since even once convergence has been reached the free energy evaluated as part of the iterations will continue to vary due to the sampling. In this work we ran all experiments for a fixed maximum number of epochs and judged convergence based on being with a fixed tolerance the final solution. For the models explored here substantial difference in the convergence of the sVB and aVB solutions was not observed, but exponential models are likely acceptably approximated by the Taylor expansion used in the aVB formulation.

Unlike the analytic solution, sVB has a number of algorithmic parameters that affect convergence that need to be chosen. These have been explored in the context of biexponential fitting. A range of parameter values were found in which convergence was robustly achieved. These were then used in an application to kinetic model fitting with success. The choice of learning rate and posterior sample size affected the best free energy achieved at convergence, although the actual variation observed in the data was relatively small. The use of a large learning rate could result in a solution near to, but not quite reaching, the global minimum due to the use of too coarse a search. Likewise, too small posterior sample size could lead the algorithm to only approach but never reach the global minimum. For small learning rates, it is possible that the algorithm takes too small a step on each iteration and either the minimum is never reached or was not reached within the maximum number of epochs allowed here. In general, very small posterior sample sizes were acceptable, as long as they were not combined with too large a learning rate, especially when there were a relatively large number of measurements in the data set. This is consistent with wider practice in the machine learning community of using a single posterior sample for the gradient calculation. This is, however, generally done in the context of far larger data than is seen in the problems explored here (i.e., a value of $N$ orders of magnitude larger than here), as demonstrated here there is greater robustness to small posterior sample size when there are more samples in the data.

The combination of limited posterior sample size combined with the use of batches, resulted in substantial reductions in the time to convergence, which is particularly helpful for applications such as ASL kinetic model fitting where a large number of series need to be analyzed representing a full volume of brain imaging data. A practical approach, but not explored here, would be to vary the learning rate and sample size during the iterative process, something that has been adopted in other machine learning applications of stochastic variational methods. This would allow a process of progressive refinement where in early stages large and 'noisy' steps are made toward the solution, with progressively smaller and more precise calculations being performed as the solution is approached.

The framework used here is similar to the variational auto-encoder popular in deep learning applications [9]. The major differences are that the decoder network has been substituted with the non-linear model, and no encoder is required, as the 'learning' process is used to obtain optimal values for the latent variables from a given dataset. Since the latent variables in this application are defined by the non-linear model, they are directly interpretable and are the target of the inference procedure. Unlike the typical application of auto-encoders, where the learning process is applied once on a large dataset to learn a representation of the data, for subsequent application to new data. Here, the framework is applied to 'learn' the parameters of a model every time it sees new data. Thus, the computational cost of 'learning' is more critical than in other machine learning applications since it determines how long it takes for new data to be processed. The results obtained here on brain imaging data suggest that sVB can be competitive in comparison to the existing aVB algorithm. In practice, the sVB framework offers various opportunities for parallelization of processing that haven't been exploited here.

# Stochastic Variational Bayesian Inference for a Nonlinear Forward Model - Supplementary Material

Michael A. Chappell, Martin Craig, Mark W. Woolrich

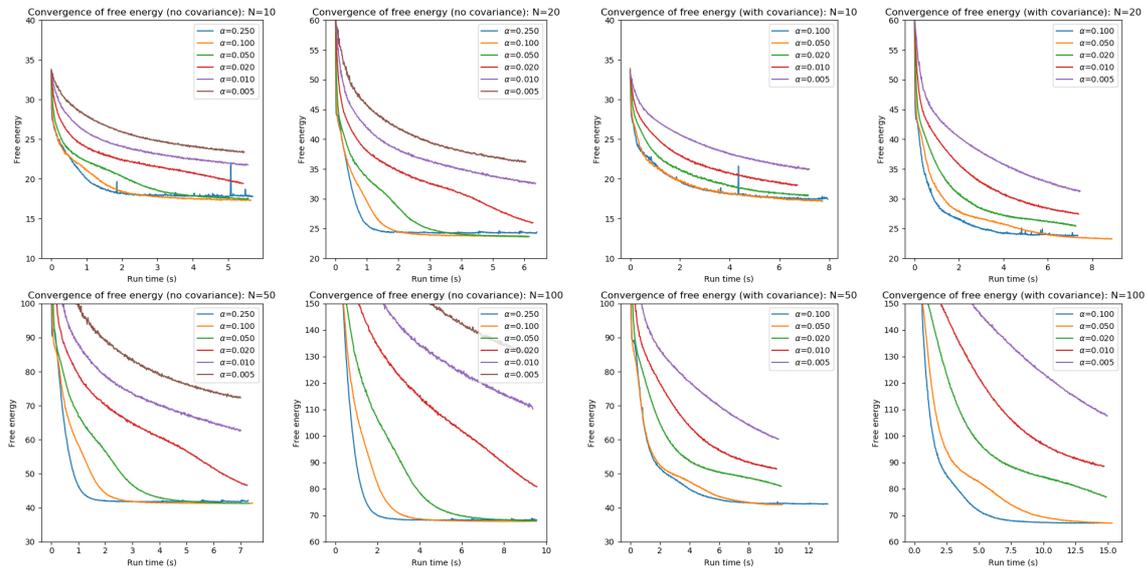

Figure S1: Convergence of free energy with runtime for different learning rates, using a sample size of 10. Faster learning rates give faster initial convergence but did not always attain the lowest free energy. Learning rates of 0.5 (and 0.25 when including covariance) are not shown as the free energy fluctuated unstably, rather than converging.

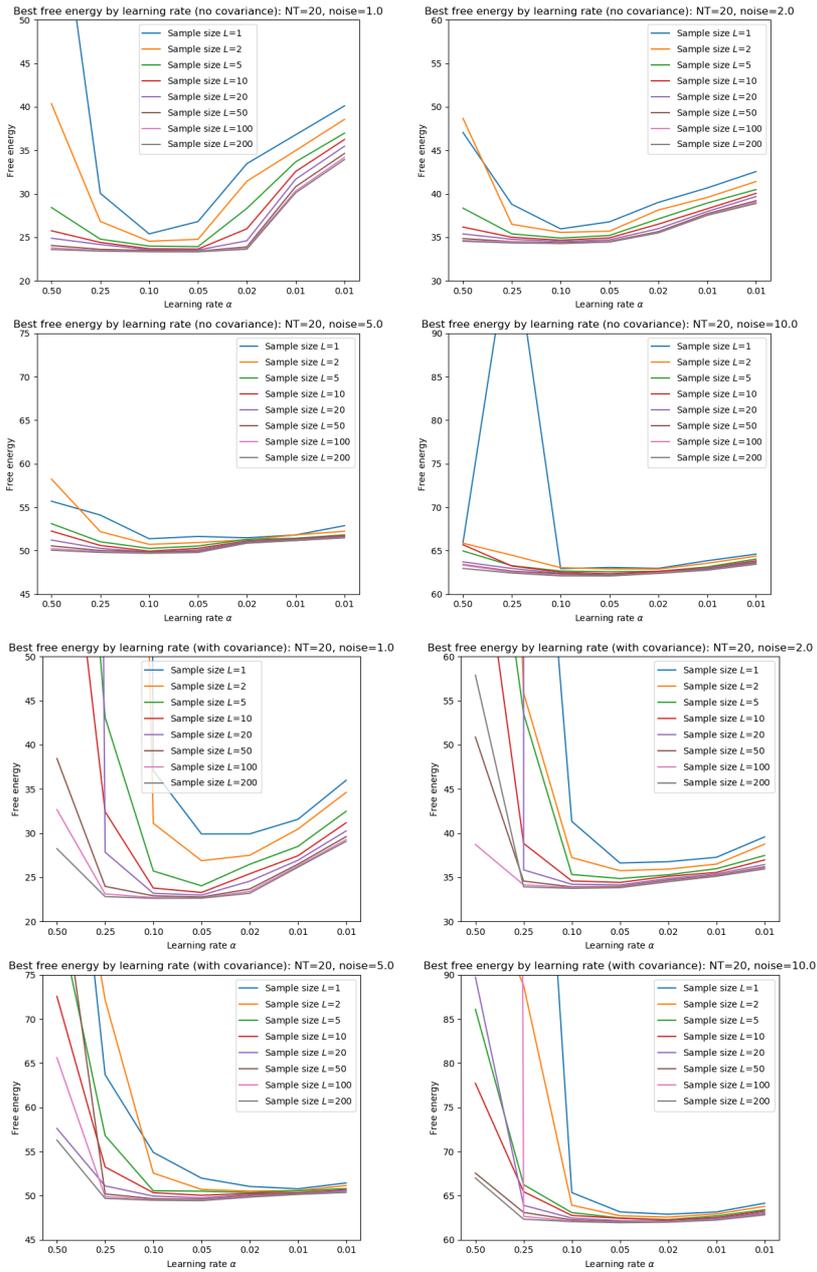

Figure S2: Mean free energy achieved in the biexponential model fitting to simulated data experiments as learning rate and number of posterior samples used was varied. Showing data size (N) of 20 for a range of different magnitude of added white noise.

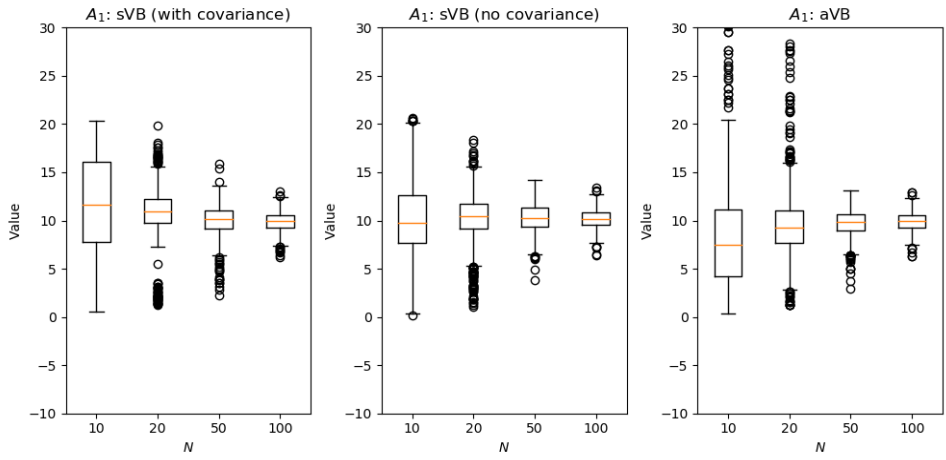
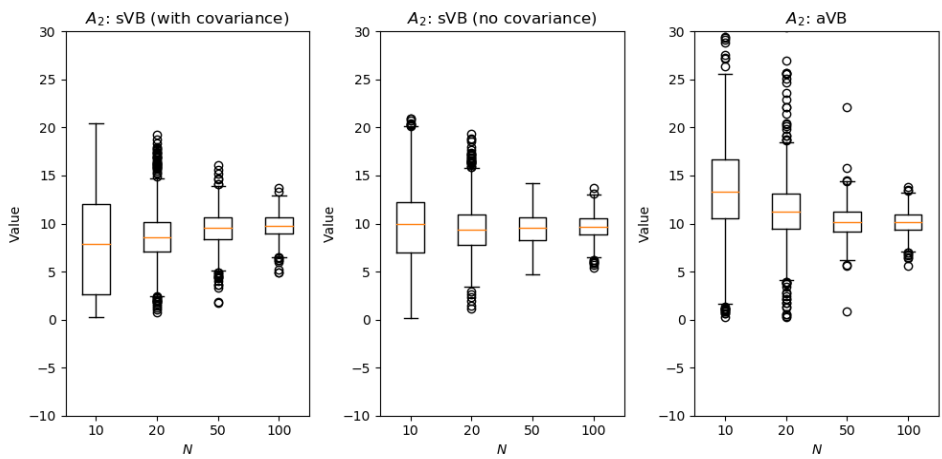
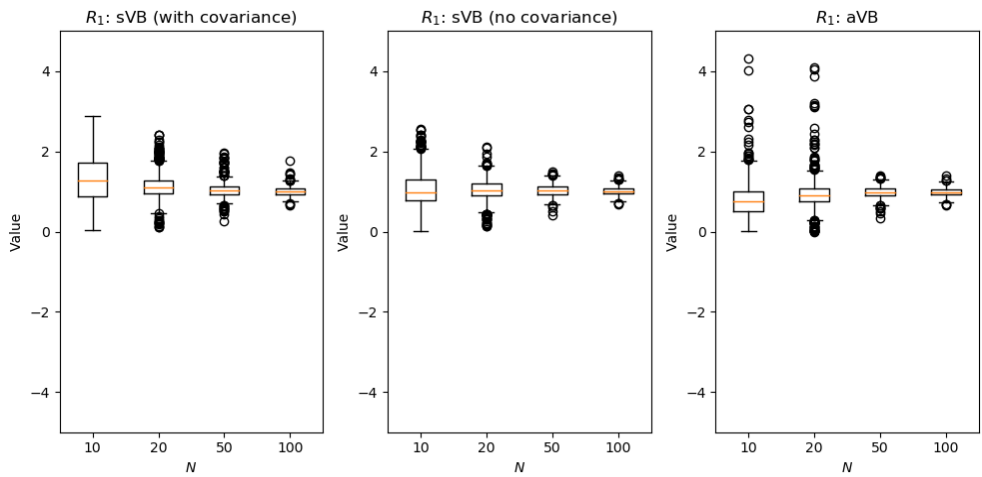

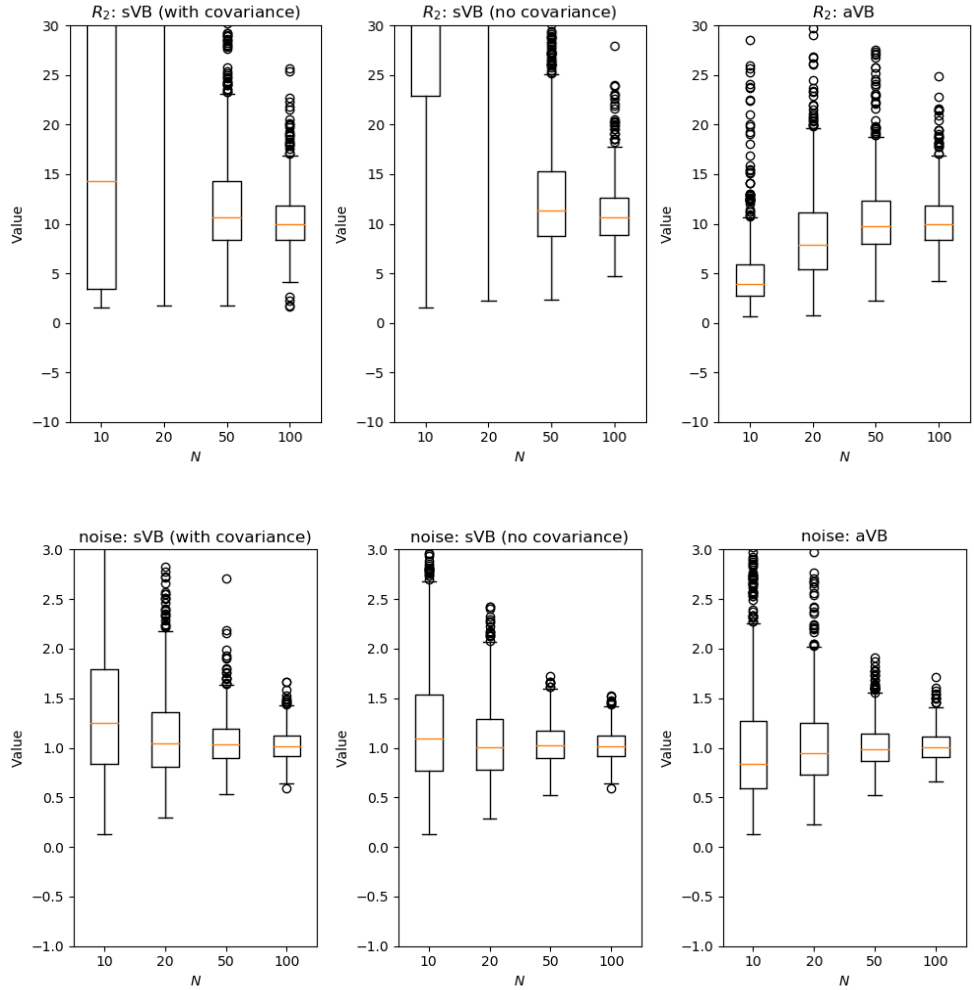

Figure S3: Parameter recovery for aVB and aVB algorithms applied to simulated data with $\alpha = 0.05$, $L = 20$ and $B = 10$. For small N (10 or 20) both aVB and sVB methods tended to estimate a single dominant exponential component. In these results this appears for aVB in the first estimates component: an increase in A1 and R1 compared to the true value. For aVB this appears in the second estimated component as an increase an A2 and a decrease in R2 compared to the true value. However, this difference is somewhat artificial and arbitrary since it depends on the sorting of the components done post-inference and doesn't necessarily represent any specific difference in the solutions found in practice.

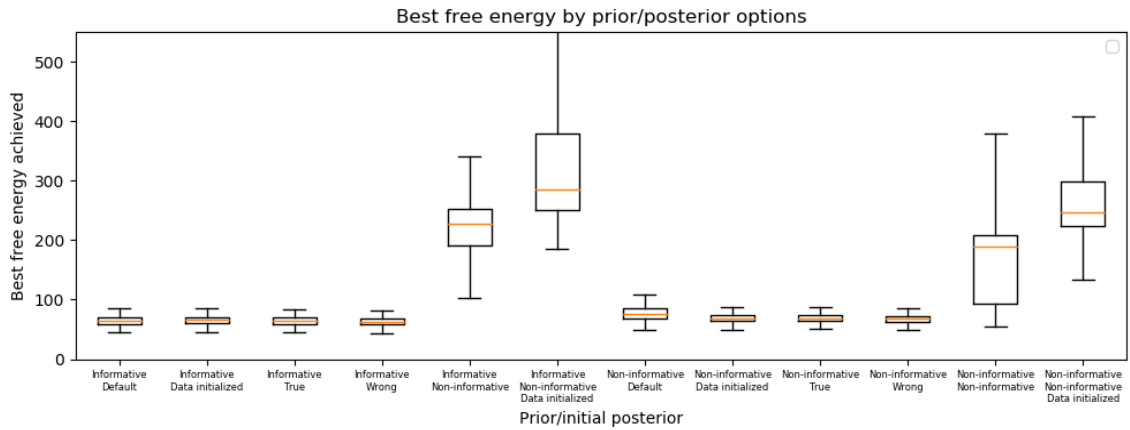

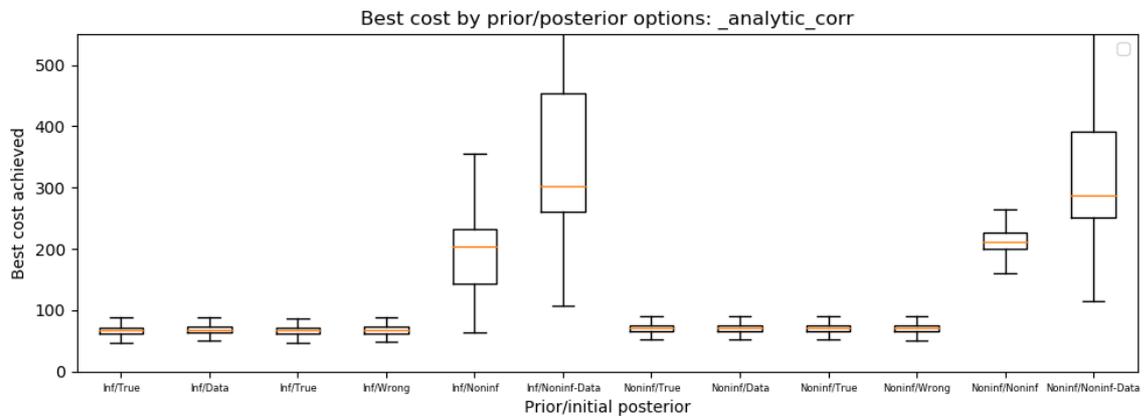

Figure S4: minimum free energy achieved for different combinations of prior and initial posterior distributions in the biexponential model fitting to simulated data experiments. This is the full set of combinations tested, Figure 5 shows a subset of these results.

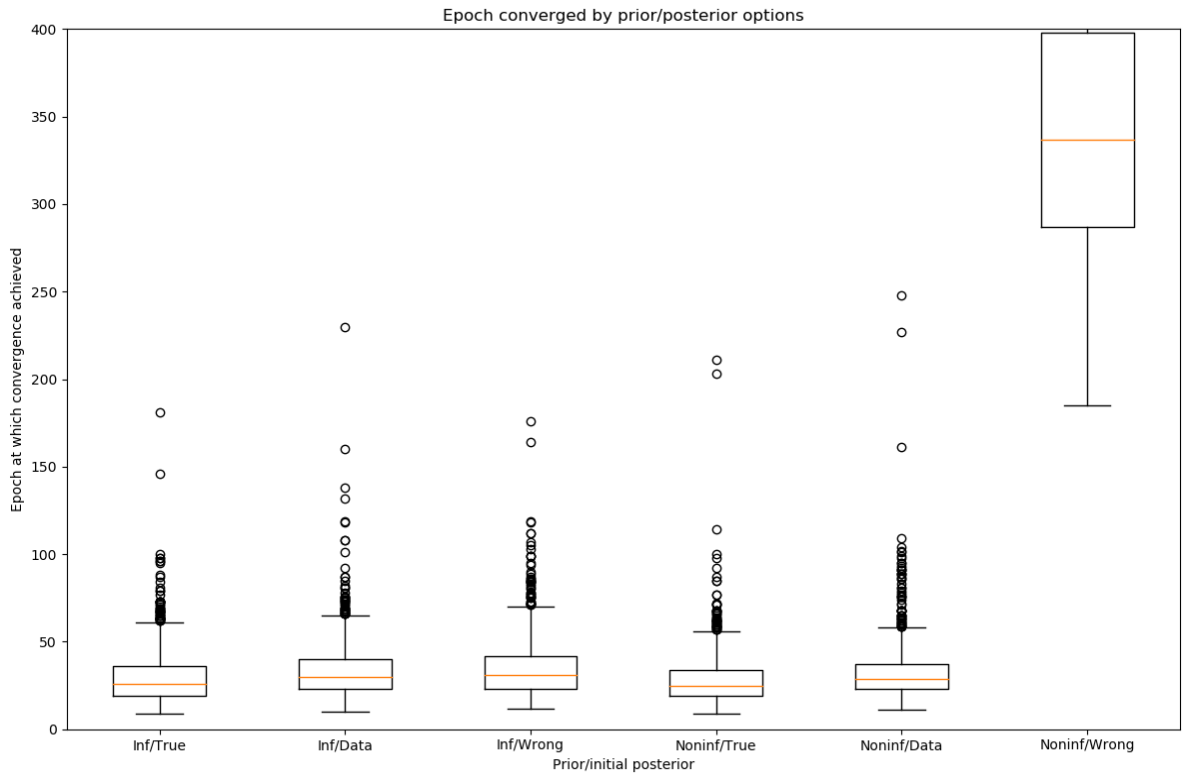

Figure S5: Number of epochs to convergence for different combinations of prior and initial posterior distribution in the biexponential model fitting to simulated data experiments.